# Dendrimer Assisted Dispersion of Carbon Nanotubes: A Molecular Dynamics Study

**Debabrata Pramanik[1] and Prabal K Maiti[1]**

[1]Centre for Condensed Matter Theory, Department of Physics, Indian Institute of Science, Bangalore 560012, India

**Abstract**

Various unique physical, chemical, mechanical and electronic properties of carbon nanotube (CNT) make it very useful materials for diverse potential application in many fields. Experimentally synthesized CNTs are generally found in bundle geometry with a mixture of different chirality and present a unique challenge to separate them. In this paper we have proposed the PAMAM dendrimer to be an ideal candidate for this separation. To estimate efficiency of the dendrimer in dispersion of CNTs from the bundle geometry, we have calculated potential of mean forces (PMF). Our PMF study of two dendrimer wrapped CNTs shows lesser binding affinity compared to the two bare CNTs. PMF study shows that the binding affinity decreases for non-protonated dendrimer and for the protonated case, the interaction is fully repulsive in nature. For both the non-protonated as well as protonated cases, the PMF increases with increasing dendrimer generations from 2 to 4 gradually compare to the bare PMF. We have performed PMF calculations with (6,5) and (6,6) chirality to study the chirality dependence of PMF. Our study shows that the PMFs between two (6,5) and two (6,6) CNT's respectively are ~ - 29 kcal/mol and ~ - 27 kcal/mol. Calculated PMF for protonated dendrimer wrapped chiral CNT's is more compared to the protonated dendrimer wrapped armchair CNTs for all the generations studied. However, for non-protonated dendrimer wrapped CNTs such chirality dependence is not very prominent. Our study suggests that the dispersion efficiency of protonated dendrimer is more compared to the non-protonated dendrimer and can be used as an effective dispersing agent in dispersion of CNT from the bundle geometry.

## 1. Introduction

Various unique properties[1,2] of carbon nanotube (CNT) make it as a potential candidate in many fields of application[3-10] in the recent times. But, the poor solubility of CNT in aqueous[11,12] and organic solutions makes dispersion and sorting of CNTs from the bundle geometry a very challenging and formidable task. To address this problem, many studies[12-20] have been performed with different polymers[16,21-25] and surfactant[13,15,26,27] molecules which work as a dispersing agent. But these methods were not so efficient in dispersion of CNTs. Recent experimental studies[28-31] have reported that DNA can be used as an effective dispersing agent in dispersion of single walled carbon nanotube (SWCNT). Several recent studies[32,33] have shown that single stranded DNA (ssDNA) and SWCNT form a stable hybrid structure in the aqueous solution and thus make dispersion of CNTs easier. Molecular dynamics (MD) studies[34-39] have reported that the hydrophobic nucleobases of ssDNA wrapped around the SWCNT surface via π-π base stacking interaction whereas the hydrophilic phosphate backbone is exposed to the aqueous environment. Thus the ssDNA-SWCNT forms a stable hybrid structure making it soluble in the aqueous medium. Earlier we have shown both experimentally and theoretically[40,41] that PAMAM dendrimer wraps SWCNTs via charge transfer interaction. This prompted us to ask the question, if this wrapping can help make SWCNT soluble and effectively disperse the nanotube. To answer this question here, we have presented a computational study of the effective interaction of the dendrimer wrapped SWNTs with PAMAM dendrimer[42-47] for different generations at various protonation levels. To calculate the effective interaction between dendrimer wrapped SWNTs as a function of dendrimer generation, we have used Umbrella sampling (US) to calculate the potential of mean force (PMF). Our calculated PMF shows that G3 and G4 protonated PAMAM dendrimer makes the SWNT completely soluble and is as efficient as ssDNA in dispersing SWNTs. To our knowledge this is the first such PMF studies for dendrimer-nanotube systems. Dendrimer also offers a distinct advantage over ssDNA because of its pH responsive wrapping mechanism and resulting inter-dendrimer repulsive interaction[48]. The rest of the paper is organized as follows: in section 2 we give the details of the system preparation along with the various computational methods used in this work. In section 3 we present our results on the PMF calculation using Umbrella sampling. Finally in section 4 we conclude with some future outlook.

## 2. Computational Details and Methodologies

We have performed all atoms Molecular Dynamics (MD) simulation using AMBER10[49] packages and ff10 force field parameters[50]. To build SWCNT, carbon atoms were modeled as the uncharged Lennard-Jones particles with sp$^2$ hybridization and ff10 force field parameter values have been used for other terms like bond-lengths, angles, dihedrals (carbon atom type in the parameter file is (CA)). Water has been modeled as the rigid molecule by using TIP3P water model[51]. Dreiding[52] force field parameters have been used to model the PAMAM dendrimer. Here, we have studied SWCNT with (6, 5) and (6,6) chirality, both having lengths of ~34 Å and diameters 7.46 Å and 8.14 Å respectively (Figure 1). We have used the ion force field developed by Joung and Cheatham[53].

To generate the dendrimer wrapped SWNT structure, first we have performed simulation of single PAMAM dendrimer and single SWCNT of (6,6) and (6,5) chirality. Initial 3-D model of the protonated and non-protonated dendrimer of various generation PAMAM dendrimer was taken from our previous work.[41] Initially PAMAM dendrimer of a given generation (either G3 or G4) was placed in close vicinity of SWCNT using xleap module of AMBER. The resulting structure was solvated in a box of TIP3P[51] water with 20 Å solvation layer in all three directions. The resulting solvated structure was minimized for 1000 steps using steepest descent method followed by 2000 steps of conjugate gradient minimization method. During the minimization, dendrimer-SWCNT was kept fixed at their initial positions by a harmonic potential with force constant of 500 kcal/mol-Å$^2$. This helps eliminate the bad contacts of water with dendrimer and SWCNT structure. The systems were then heated gradually in the NVT molecular dynamics starting from 0 K to 300 K by using a weak harmonic restrain of 20 kcal/mol-Å$^2$ on the dendrimer-SWCNT structure. To maintain the system at a constant temperature and pressure, NPT MD (P = 1 atm, T = 300 K) was performed for 120 ps using Berendsen weak coupling method[54] with a coupling constants of 0.5 ps both for the temperature and the pressure bath coupling. The Particle Mesh Ewald (PME)[55] method was used to calculate the long range electrostatic interactions using 4$^{th}$ order cubic B-spline interpolation with a tolerance of 10$^{-4}$. For both the long range electrostatic interaction as well as short range van der Waals interaction, a real space cutoff of 9 Å was used. The non-bonded list was updated after every 10 steps. The SHAKE algorithm[56] was used to constraint the bonds involving hydrogen atoms and 2 fs time step was used. Finally 30 ns long production run was performed in NVT ensemble. During 30 ns long production run, dendrimer got adsorbed onto the nanotube surface and produced stable dendrimer-nanotube composite structure as shown in our previous study[40,41]. To calculate the PMF[57,58], we have taken two copies of this composite structure and placed them parallel to each other keeping a distance of 30 Å between their center of mass (COM) (Figure 1). This system was solvated by adding a 20 Å solvation layer in all three directions from the dendrimer-SWCNT composite. In addition, some water molecules were replaced by the Cl- counter ions to neutralize the positive charges of the protonated dendrimer. Periodic boundary condition (PBC) was used for all the cases in all three directions. Details of the system used for the PMF calculations are given in Table 1. These systems are then subject to similar simulation protocol as described above before starting the Umbrella sampling simulations.

PMF was calculated by using Umbrella Sampling (US) technique[59] using a harmonic potential with force constant of 4 kcal/mol-Å$^2$. The COM distance between two SWCNTs was taken as the reaction coordinate. To calculate PMF between two dendrimer-SWCNT composite structures, one SWCNT was kept fixed while changing the COM distance of the second SWCNT. By fixing one SWCNT, the other SWCNT was brought closer by gradually decreasing the reaction coordinate up to 5 Å starting from 30 Å (Figure 1). Finally using WHAM (Weighted Histogram Analysis Method) technique[60-62], we calculated PMF for the unbiased system by subtracting the contribution of the additional harmonic potential. To calculate PMF between SWCNT-dendrimer using US, a force constant of 5 Kcal/mol-Å$^2$ was used. Here, our earlier equilibrated nanotube-dendrimer composite structure was taken as the initial configuration for the PMF calculation. Using center to center distance as the reaction coordinate between SWCNT and dendrimer complex, the dendrimer molecule was gradually brought away from the nanotube surface starting from 0 Å to 30 Å with 1 Å window. At each window 2 ns production run was performed to collect data for PMF calculation. Next, PMF was calculated the same way as described above for other composite systems by using WHAM. We have done at least three independent PMF calculations using different initial configurations for all the cases presented in this manuscript.

3. Results and Discussions

**PMF between single dendrimer and bare SWNT**

Earlier studies[40,41] have reported that the dendrimer forms a stable complex with bare SWNT. A detailed description about various properties of this dendrimer-SWNT composite structure has been reported in our earlier studies. Here our interest was to know how a single dendrimer interact with bare SWNT at room temperature at 300 K. To answer this, we have carried out PMF calculation for 2$^{nd}$ generation protonated PAMAM dendrimer and chiral nanotube (6,5) by using US technique. The PMF profile is shown in Figure 2. The PMF profile shows a very strong attraction of ~ -110 kcal/mol between dendrimer and SWCNT at a COM separation of 3 Å between the two. Below this distance, nanotube and dendrimer repel each other due to excluded volume interaction. Away from this distance, attraction between nanotube and dendrimer decreases resulting in increase in PMF up to a distance of ~ 30 Å. After this distance, the dendrimer goes away from the interaction region of the carbon nanotube. Figure 2 shows 3 different instantaneous snapshots of the nanotube-dendrimer at three different distances. The instantaneous molecular snapshots provide a microscopic level picture of interaction between dendrimer and SWCNT as dendrimer goes away from the nanotube surface.

To disperse SWNT from the bundle geometry, it is important to know the strength of the interaction between SWNTs. To get a quantitative as well as microscopic picture of interaction, we have carried out PMF calculation between two bare SWNTs using US technique.

**PMF between two bare SWNTs (6,5)**

We have calculated the PMF between two bare chiral (6,5) SWNTs using the same protocol as mentioned in the above section. To calculate PMF, each window was taken of 1 Å size and at each window, the system was simulated for 1 ns. The PMF profile is shown in Figure 3. It shows that two SWNTs are non-interacting at around 30 Å distance and then with decreasing inter-tube distances, nanotubes start interacting with each other by van der Waals potential. The PMF is minimum at ~ 11 Å center to center separations and it gradually increases away from this distance. The minimum value of PMF is ~ – 29 kcal/mol. We also show the snapshots of the system showing the orientation of the tubes at various inter-tube separations. At the minimum of PMF the two bare nanotubes are parallel and bind with each other very strongly by strong van der Waals attraction. This strong binding between two bare nanotubes is of the order of 29 kcal/mol which is very high compared to the thermal energy of ~ 0.59 kcal/mol at room temperature at 300 K. Because of this, only thermal energy is not sufficient to disperse nanotube from the bundle geometry by overcoming such strong attractive interaction. Due to this reason, dispersion and separation of pristine carbon nanotube from bundle geometry is very challenging task.

To address this problem we need to introduce some external agents which can wrap onto the nanotube surface and make it soluble in the aqueous environment and which will be able to screen the large van der Waals attraction between the two nanotubes. From our earlier studies, we were motivated to carry out studies on dendrimer as an external agent. How the interaction between two SWNTs modifies in presence of the dendrimer is demonstrated here below.

### Dendrimer as external agent

Here we have studied the PMF of the dendrimer wrapped CNTs with PAMAM dendrimer at different protonation levels (non-protonated (NP) and protonated (PP)) and for three different generations, G2, G3 and G4. The systems were prepared using the same simulation protocol as mentioned in Section 2. Initially the two dendrimer wrapped CNTs were placed at 30 Å apart and then gradually their distances were reduced in steps of 1 Å (window size) until the inter-tube separation reduced to 5 Å. At each window the system was simulated for 1 ns. The calculated PMF profiles for the case of G2 dendrimer (both PP and NP) are shown in Figure 4. For comparison, we have also shown PMF profile for bare (6,5) SWNT in Figure 4. We see that PMF between dendrimer wrapped (6,5) SWNTs increases (both for PP and NP dendrimer) compared to the PMF between two bare (6,5) SWCNTs. For example, the minimum of the averaged PMF for NP G2 dendrimer wrapped SWNT is ~ - 4 kcal/mol, compared to the ~ - 29 kcal/mol between two bare SWNTs. For the PP G2 dendrimer wrapped SWNT we find no evidence of attractive interaction between the SWNTs and they become completely repulsive. Thus both the NP and PP G2 dendrimer when added as external agent, wrap around the SWNT and screen the strong van der Waals attraction between the two nanotubes. The screening decreases the binding affinity between two bare nanotubes. In the case of PP dendrimer, the effective interaction between SWNTs becomes completely repulsive. Thus, dendrimer emerges out to be a good agent to lower the inter-binding affinity between two bare SWNTs.

In our earlier study,[41] we have reported that the number of close contacts of dendrimer-SWNT composite increases with increasing dendrimer generations both for the PP and NP dendrimers. So, the wrapped surface coverage of the dendrimer onto the nanotube surface also increases with increasing generations. So the effective interaction between the dendrimer wrapped SWNT should depends on the dendrimer generations as well. To probe the effect of dendrimer generations, we have also calculated PMF between dendrimer wrapped SWNTs for higher generation dendrimers like G3 and G4 for both the PP and NP cases as well.

**Dependency of PMF on dendrimer generations**

The averaged PMF profiles for NP dendrimer wrapped SWNT for generations G2, G3 and G4 are shown in Figure 5. For comparison, we have also shown the PMF profile for two bare SWNTs. We find that with increasing dendrimer generations, the PMF increases compared to the bare SWNTs. For example, for bare SWNTs, the PMF is - 29 kcal/mol, which increases to ~ - 4 kcal/mol for G2 wrapped SWNTs. The PMF shows completely repulsive behavior for G3 and G4 wrapped SWNTs. So if one prefers to use NP dendrimer, one can use higher generation dendrimers like G3, G4 for dispersing the SWNTs. The PMF profiles for the case of PP dendrimer wrapped SWNTs for various generations are shown in Figure 6. The profiles show that in case of PP dendrimers, PMFs for all the three generations are positive and hence SWNTs do not attract each other. For each case we have performed three independent PMF calculations and showed an average PMF with a representative error bar. From figure 6 we find that with increasing dendrimer generations the slope of the PMF gradually increases in going from G2 to G3. For G3 and G4, the PMF profiles show almost similar behavior. It might be due to the fact that, G3 PP dendrimer is sufficient to cover the whole length of the nanotube surface and so surface coverage does not change in going from G3 to G4. So, it shows that for PP dendrimer, dispersion efficiency increases with increasing generations.

As the PP dendrimer is overall positively charged, the electrostatic interaction plays an important role to screen the attractive interaction arising due to strong van der Waals interaction between the carbons of SWNTs. For this reason, PP dendrimer turns out to be very effective in dispersion of bare nanotubes compared to the NP dendrimer. The PMF profiles shown in Figure 7 give a comparison of the PMF between NP and PP dendrimer wrapped SWNTs for various generations. It shows that for SWNTs wrapped with G2 NP dendrimer, PMF profile shows an effective attraction (~ - 4 kcal/mol) whereas for SWNTs wrapped with G3 and G4 NP dendrimers or PP dendrimers of generation 2 to 4, PMFs are positive and so repulsive in nature.

**Chirality dependence of PMF**

The wrapping of dendrimer depends on the geometry of the nanotube surface. The adsorption of dendrimer onto the nanotube surface changes with the chirality of the SWNT. So it is interesting to know how the PMF between two dendrimer wrapped SWNTs vary with the chirality of the SWNT. Here we have carried out PMF studies with two types of chirality, (6,5) and (6,6) respectively having similar diameter. We have calculated PMF for both the chirality, wrapped with NP as well as PP dendrimer for 3 different generations (G2 to G4). Figure 8 (a) compares the PMF between two bare (6,5) SWNTs and (6,6) SWNTs. The minimum of the PMF for two bare (6,5) SWNTs is ~ - 29 kcal/mol compared to ~ - 27 kcal/mol for two bare (6,6) SWNTs. High negative PMF signifies that there exist very strong binding affinity between two bare SWNTs for both the chiral types. The binding affinity for (6,5) SWNT is higher than (6,6) SWNT by ~ 2 kcal/mol. The strength of the binding between two nanotubes depends on the stacking of nanotubes onto one another. As the stacking varies with chirality of the nanotube, so it's expected to have different binding affinity for (6,5) and (6,6) SWNTs. As (6,5) SWNT has better stacking conformation compared to (6,6) nanotube, it gives rise to the higher binding affinity for (6,5) nanotube in comparison to the (6,6). Similar trend is also reflected for the dendrimer wrapped SWNTs for NP case (Figure 8 (b)). For PP dendrimer wrapped SWNTs, there is no such trend as shown in Figure 8 (c). Figure 8 (b) shows that for NP dendrimer wrapped SWNTs, the change in PMF for armchair (6,6) CNT is more compared to chiral (6,5) CNT for all the three generations. As the binding affinity for two (6,5) SWNTs is more compared to that of two (6,6) SWNTs, it is consistent that when NP dendrimer wraps SWNTs, the reduction in binding strength for armchair will be more compared to chiral CNT. For both types of chirality, the PMF values increase with increasing generations going from G2 to G4. Figure 8 (b) shows that PMF for (6,6) CNT is repulsive whereas PMF for (6,5) CNT is attractive when wrapped by G2 NP dendrimer. For G3 and G4 NP dendrimer wrapped SWNTs, PMFs are repulsive for both the types of chirality. Thus we conclude that G3 and G4 NP dendrimers can be used as an effective dispersing agent to disperse SWNT of specific chirality from the bundle geometry. For PP dendrimer of different generations, the PMF profiles between two dendrimer wrapped SWNTs, for both (6,5) and (6,6) chirality, are almost similar in nature as shown in Figure 8 (c). As PP dendrimers are charged molecules, the interaction between two nanotubes is mostly modulated by the PP dendrimer interaction in between two CNTs and chirality doesn't play much role. In case of protonated dendrimer, for all the three generations and for both the types of SWNTs, the PMF values are fully positive. So, the interactions are fully repulsive in nature in the presence of the PP dendrimer. Thus PP dendrimers can be used as very effective dispersing external agent to disperse SWNT from the bundle geometry.

**Conclusions**

To summarize, using all atom MD simulation and US technique, in this paper we compute the PMF between two bare CNTs in aqueous environment as a function of tube-tube separation for (6,5) and (6,6) chirality. Our calculation suggests that bare CNTs are strongly attractive and only thermal energy is not sufficient to disperse nanotube to overcome this strong attraction. Our calculated PMFs show that introduction of the dendrimer as an external agent between two nanotubes screen the strong inter nanotube attraction. Calculated PMFs show that for NP G2 dendrimer, PMF increases and for NP G3 and G4 dendrimer, the PMF becomes positive. For PP dendrimer, PMFs become positive for all three generations G2, G3 and G4. We have calculated PMF with (6,5) and (6,6) chirality to study the chirality dependence of PMF. When PMF is calculated with dendrimer wrapped SWNT, we find that the changes in PMF for armchair SWNT is more compared to the chiral SWNT for NP dendrimer for all the generations of dendrimer from 2 to 4. For PP dendrimers we see, PMFs for all cases are repulsive in nature. So our PMF study shows that the dispersion efficiency of PP dendrimer is more compared to the NP dendrimer for all three generations. For NP dendrimer, PMF increases for G2 dendrimer and for G3 and G4, the PMF becomes fully positive. So, PP dendrimer of G2-G4 and NP G3 and G4 dendrimers can be used as an effective dispersing agent in dispersion of SWNT from the bundle geometry. It is worth mentioning here that all our conclusions are based on effective two body PMF between dendrimers and dendrimers wrapped CNTs and questions remain whether many body interactions[63] in such soft colloid systems can be significant. In the context of dendrimer interaction, many body effects have not been investigated well. Terao[64] reported both the 2-body and three-body interaction using a coarse-grained MD simulation of dendrimer. He demonstrated that both the two-body and three body forces decay as a function of distance and are of similar nature. The triplet force is repulsive in nature. However, their calculation did not include the presence of explicit solvent. However, previous small angle neutron scattering studies (SANS) along with the theoretical calculations[65,66] have demonstrated that for dilute solution of dendrimers, effective two body interaction can fit the experimental dendrimer-dendrimer structure factor very well. In fact Likos et. al. have shown that[66] for a range of dendrimer concentrations, two body Gaussian effective interaction fits the experimentally observed structure factor very well. So we believe our two body PMF between dendrimer wrapped CNTs is well justified for the problem of CNT separation. Earlier we have also shown[48] that the effective interaction between dendrimer can be well represented by two body interaction using a sum of exponential and Gaussian function. So our future work will involve calculating many body interactions in dendrimer solution as well as dendrimer wrapped CNT systems at the atomistic level to better understand their significance.

## Acknowledgements

We thank Supercomputer Education and Research Centre, IISc for providing supercomputer facilities. We thank DST, India for financial support.

# Tables and Figures

**Table 1: Details of the simulated systems reported in this work. The table gives information regarding the number of water molecules, counter ions, total number of atoms, box dimensions, reaction coordinates and number of US windows for all the simulations performed.**

| Systems (PMF performed) | No. of water molecules | No. of counter-ions (Cl-) | Total no. of atoms | Box dimensions ($Å^3$) | Reaction coordinates (Å) (Initial, Final) | No. of US window |
|---|---|---|---|---|---|---|
| CNT(6,5) and CNT(6,5) | 34404 | 0 | 35132 | 87 x 51 x 84 | 30, 5 | 26 |
| CNT(6,6) and CNT(6,6) | 32481 | 0 | 33201 | 88 x 52 x 79 | 30, 5 | 26 |
| CNT(6,5) and G2 NP | 25080 | 0 | 26840 | 73 x 54 x 86 | 30, 5 | 26 |
| CNT(6,5) and G3 NP | 92370 | 0 | 95282 | 104 x 104 x 102 | 30, 5 | 26 |
| CNT(6,5) and G4 NP | 153288 | 0 | 158504 | 134 x 119 x 113 | 30, 5 | 26 |
| CNT(6,5) and G2 PP | 66549 | 32 | 68373 | 90 x 116 x 77 | 30, 5 | 26 |
| CNT(6,5) and G3 PP | 120711 | 64 | 123751 | 93 x 136 x 114 | 30, 5 | 26 |
| CNT(6,5) and G4 PP | 162348 | 128 | 167820 | 123 x 121 x 131 | 30, 5 | 26 |
| CNT(6,6) and G2 NP | 56265 | 0 | 58017 | 110 x 77 x 82 | 30, 5 | 26 |
| CNT(6,6) and G3 NP | 81024 | 0 | 83928 | 130 x 81 x 94 | 30, 5 | 26 |
| CNT(6,6) and G4 NP | 135957 | 0 | 141165 | 127 x 132 x 98 | 30, 5 | 26 |
| CNT(6,6) and G2 PP | 54270 | 32 | 56086 | 74 x 105 x 88 | 30, 5 | 26 |
| CNT(6,6) and G3 PP | 97500 | 64 | 100532 | 93 x 92 x 138 | 30, 5 | 26 |
| CNT(6,6) and G4 PP | 179064 | 128 | 184528 | 110 x 164 x 119 | 30, 5 | 26 |

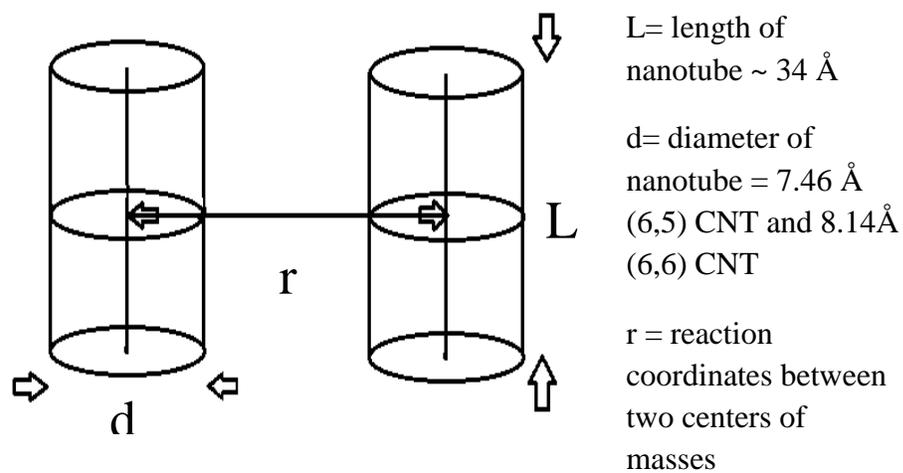

**Figure 1:** Schematic of two nanotubes and their reaction coordinates. The two nanotubes were pulled apart starting from 30 Å to 5 Å for PMF calculation.

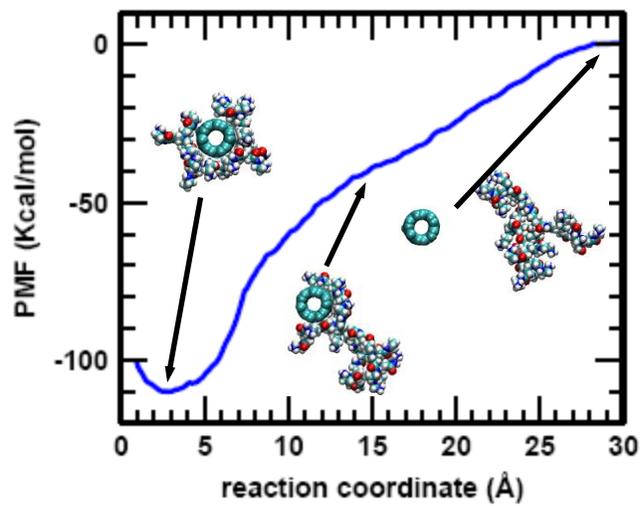

**Figure 2**: PMF as a function of the distance between the center of masses of the dendrimer and CNT (reaction coordinate). The PMF is between bare CNT (6,5) and G2 PP PAMAM dendrimer.

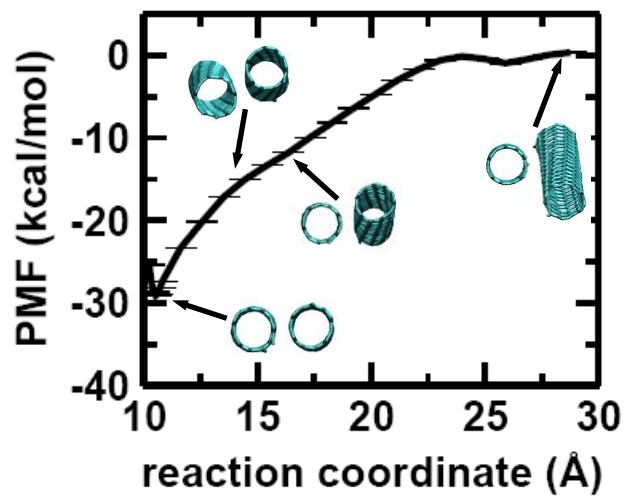

**Figure 3**: PMF as a function of the inter-tube separation. The plot shown is PMF between two bare CNTs of chirality (6,5). The instantaneous snapshots show the orientation of the CNTs at inter-tube separation distances of 10.7 Å, 13.6 Å, 16.4 Å and 27.4 Å respectively.

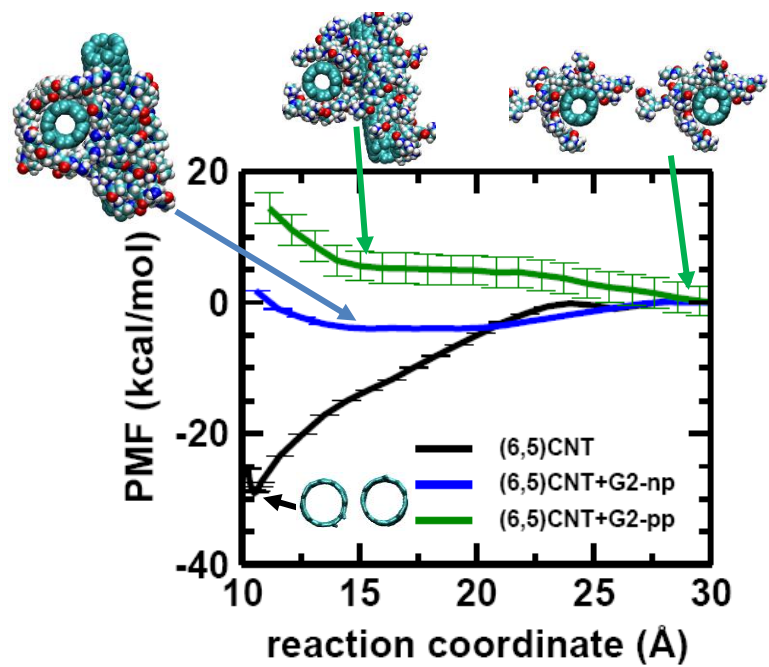

**Figure 4**: Comparison of the PMF between bare CNTs and dendrimer (PP and NP respectively) wrapped CNTs. The minima of the PMF decreases drastically for NP dendrimer CNTs compared to the PMF of bare CNTs (-29 kcal/mol for bare CNTs vs - 4 kcal/mol for NP dendrimer wrapped CNTs). The PMF becomes fully positive for PP dendrimer wrapped CNTs signifying repulsive nature of interaction between the CNTs. For both G2 NP and PP dendrimers, we have presented average PMF with representative error bar obtained from three independent PMF calculations. The snapshots show the microscopic pictures of the dendrimer wrapped CNTs at various positions of the PMF profile.

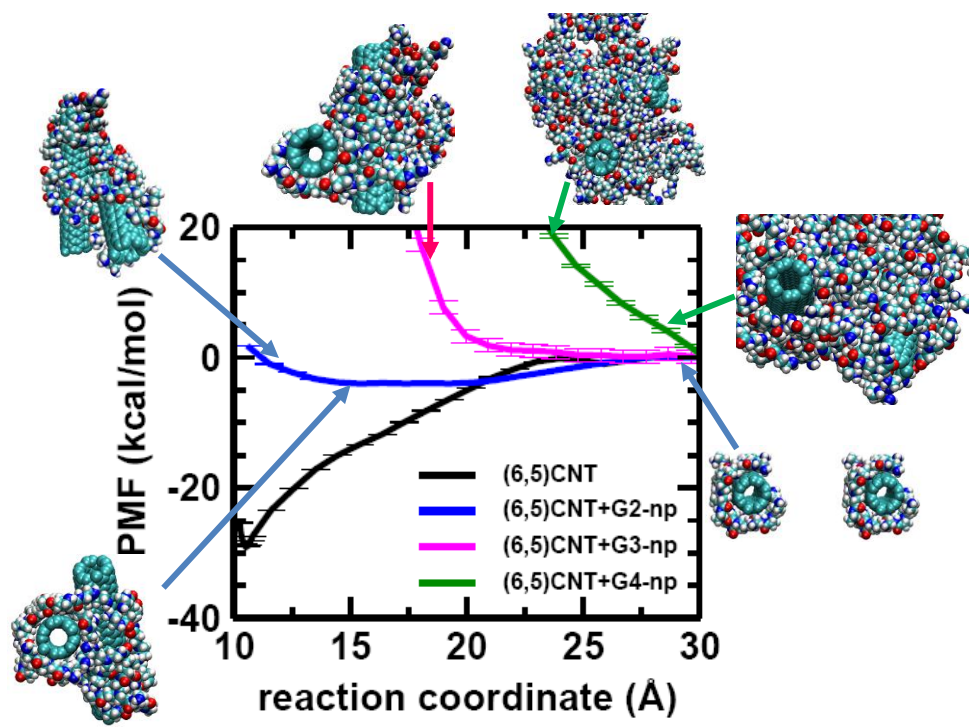

**Figure 5**: The PMF as a function of inter-tube separation for dendrimer wrapped CNTs for different generations of non-protonated dendrimer. For comparison, PMF for bare CNTs is also shown in the same plot. With the increase in dendrimer generation the minima of the PMF decreases and eventually PMF become fully positive for G4 dendrimer. For NP dendrimer wrapped CNTs, PMF is averaged over three independent PMF calculations. The instantaneous snapshots present microscopic pictures at different instant of positions along the PMF profile.

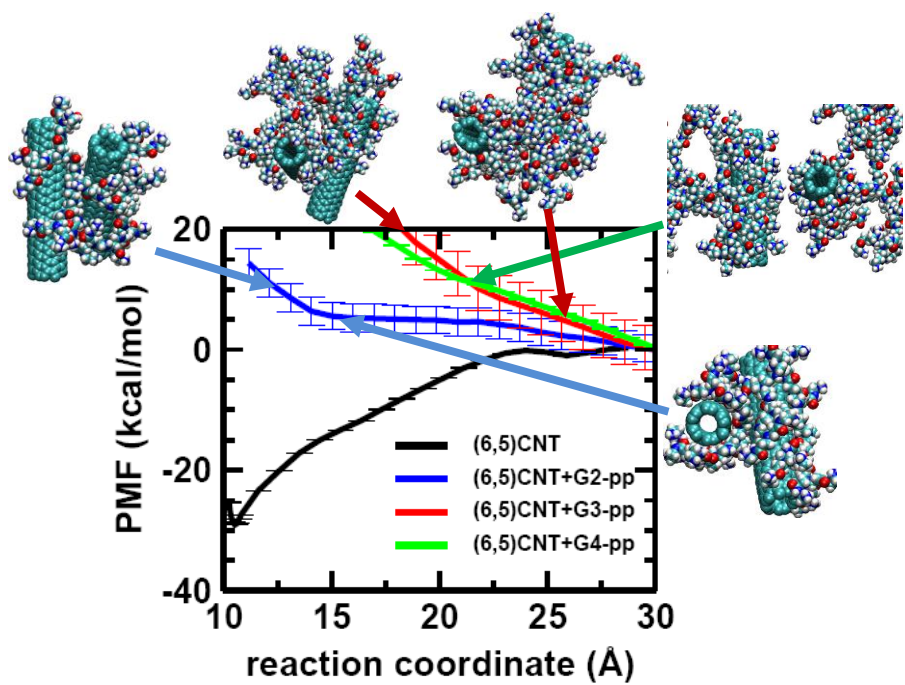

**Figure 6**: PMF profile for the PP dendrimer wrapped CNTs as a function of dendrimer generations. Wrapping of PP dendrimer makes the PMF profile fully positive for all the generations. For G2 to G4 PP dendrimers, we have put representative error bar (average of three independent PMF calculations). The instantaneous snapshots show microscopic pictures at different inter-tube separation for different dendrimer generations.

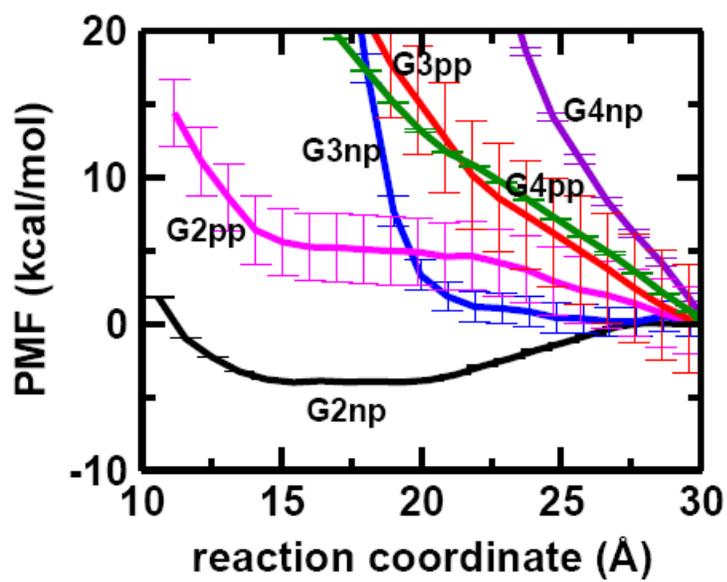

**Figure 7**: Comparison of the PMF profiles for dendrimer wrapped CNTs (6,5) for various dendrimer generations both for the NP and PP dendrimer. We see that for NP dendrimer, with increase in dendrimer generations, the PMF profiles become more and more positive.

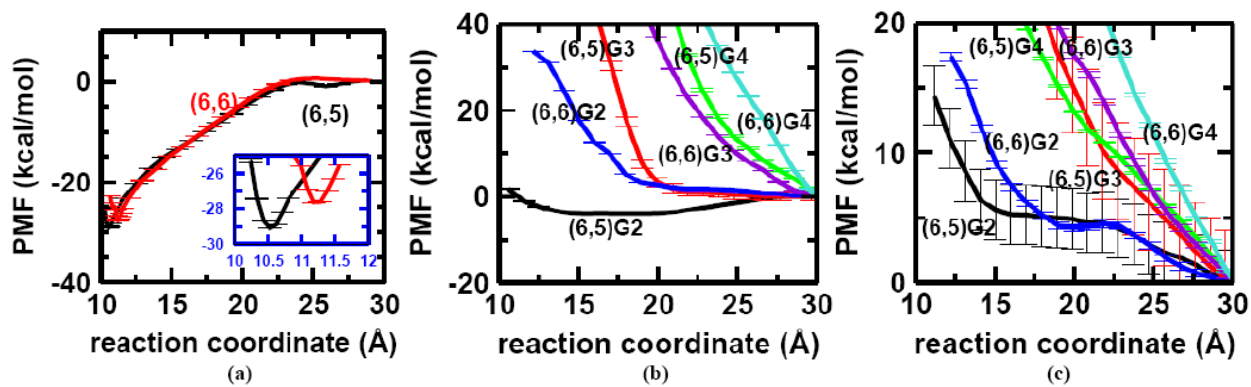

**Figure 8**: Chirality dependence of the PMF profile. (a) Comparison of PMF plot for two bare CNTs of (6,5) and (6,6) chirality. (b) PMF as a function of inter-tube distance for NP dendrimer wrapped CNTs of (6,5) and (6,6) chirality (c) PMF for PP dendrimer wrapped CNTs of (6,5) and (6,6) chirality.